\documentclass[aps,prb,twocolumn]{revtex4}
\usepackage{amsmath, amssymb, graphics, setspace}
\usepackage{amsmath}
\usepackage{graphicx}
\usepackage{dsfont}
\usepackage{multirow}

\begin{document}

\title{Inter-valley spiral order in the Mott insulating state of a
heterostructure of trilayer graphene-boron nitride}
\author{Guo-Yi Zhu$^{1}$, Tao Xiang$^{2,3}$, and Guang-Ming Zhang$^{1,3}$}
\affiliation{$^{1}$State Key Laboratory of Low-Dimensional Quantum Physics and Department
of Physics, Tsinghua University, Beijing 100084, China \\
$^{2}$Institute of Physics, Chinese Academy of Sciences, Beijing 100190,
China \\
$^{3}$Collaborative Innovation Center of Quantum Matter, Beijing, 100084, China}
\date{\today}

\begin{abstract}
Recent experiment has shown that the ABC-stacked trilayer graphene-boron
nitride Moire super-lattice at half-filling is a Mott insulator. Based on
symmetry analysis and effective band structure calculation, we propose a
valley-contrasting chiral tight-binding model with local Coulomb interaction
to describe this Moire super-lattice system. By matching the positions of
van Hove points in the low-energy effective bands, the valley-contrasting
staggered flux per triangle is determined around $\pi /2$. When the valence
band is half-filled, the Fermi surfaces are found to be perfectly nested
between the two valleys. Such an effect can induce an inter-valley spiral
order with a gap in the charge excitations, indicating that the Mott
insulating behavior observed in the trilayer graphene-boron nitride Moire
super-lattice results predominantly from the inter-valley scattering.
\newline

\end{abstract}

\maketitle

\section{Introduction}

The Moire super-lattice in the van der Waals heterostructure composed of
multi-layer graphenes and hexagonal boron nitrides (hBN) has recently
attracted great interest \cite{Yankowitz,Dean,Hunt,Ponomarenko,YangZhang,ShiWang}. Both graphenes and hBN
have hexagonal lattice structures, but the original lattice periodicity is
ruined due to the mismatch between their lattice constants. Nevertheless,
the periodicity can be restored on a much larger length scale, i.e., the
Moire wave length ($L_{\rm M}\simeq 15$ nm), upon which a triangular Moire
super-lattice emerges \cite{Dean,Hunt,Ponomarenko}. On the other hand,
bilayer graphene with a small twisted angle can also form the Moire band
structure \cite{Neto,MacDonald2011,Magaud2012,Cao2016PRL}. In the
magic-angle twisted bilayer graphene, the Moire bandwidth is reduced
dramatically and the local Coulomb repulsion becomes relatively significant,
leading to the observation of the Mott insulating state as well as the
unconventional superconductivity around the half-filling \cite{Cao2018Corr,Cao2018SC}. Meanwhile, it has been reported that a Mott
insulating state also exists in the ABC-stacked trilayer graphene-hBN
heterostructure \cite{FengWang}. In this experiment the low energy bandwidth
is about $10$ meV while a Mott gap $\sim $ $2$ meV is observed at half
filling. The comparable energy scale renders such a system in an
intermediate coupling regime, therefore the role of band structure cannot be
overemphasized.

In this paper, we will investigate the physical origin of the Mott
insulating behavior observed in the trilayer-graphene-hBN heterostructure.
Based on the symmetry analysis and effective band structure calculation, we
propose a minimal tight-binding model with local Coulomb interaction. This
model defined on a triangular lattice characterizes an interacting electron
system in a staggered fictitious magnetic field for each of the two
degenerate valley degrees of freedom. By matching the van Hove point
positions of the effective low-energy bands, the staggered flux of each
triangle is close to $\pi /2$. At half-filling, the two valley Fermi
surfaces are found to be perfectly nested. Such an effect leads to a novel
correlated insulating state with an inter-valley spiral order and a charge
excitation gap, giving a natural explanation to the experimental observation.

\section{Moire band structure}

The ABC-stacked trilayer graphene (TLG) has the same Bravais
lattice as in the monolayer graphene. But the electron and hole touching at
zero energy support chiral quasiparticles with $3\pi $ Berry phase,
generalizing the low-energy band structure of the monolayer and bilayer
graphene \cite{KoshinoMcCann}. The hBN also forms a honeycomb lattice but
has a lattice constant about $1.8\%$ larger than that of the graphene. Thus
the heterostructure of TLG and hBN can form a triangular Moire super-lattice
shown in Fig. 1a, which contains three interlaced regions. The region shaded
by blue circles shows the maximal alignment between the TLG and hBN, denoted
as the $\alpha $ zone; and the regions shaded by yellow or green triangles
have a larger misalignment between the TLG and hBN, denoted as $\beta $ and $%
\beta ^{\prime }$ zone, respectively. The $\beta $ zone differs from the $%
\beta ^{\prime }$ zone by a sub-lattice exchange, defined by the $C_{6}$
rotation along the $z$-axis or the $M_{y}$ mirror reflection with respect to
the $x$-$z$ plane. Each unit cell of the Moire super-lattice includes the $%
\alpha $, $\beta $ and $\beta ^{\prime }$ zone. The TLG-hBN heterostructure
possesses the three-fold rotational symmetry along the $z$-axis $C_{3}$, the
mirror reflection symmetry with respect to the $y$-$z$ plane $M_{x}$, and the
time reversal symmetry $\mathcal{T}$.
\begin{figure}[t]
\centering
\includegraphics[width=8cm]{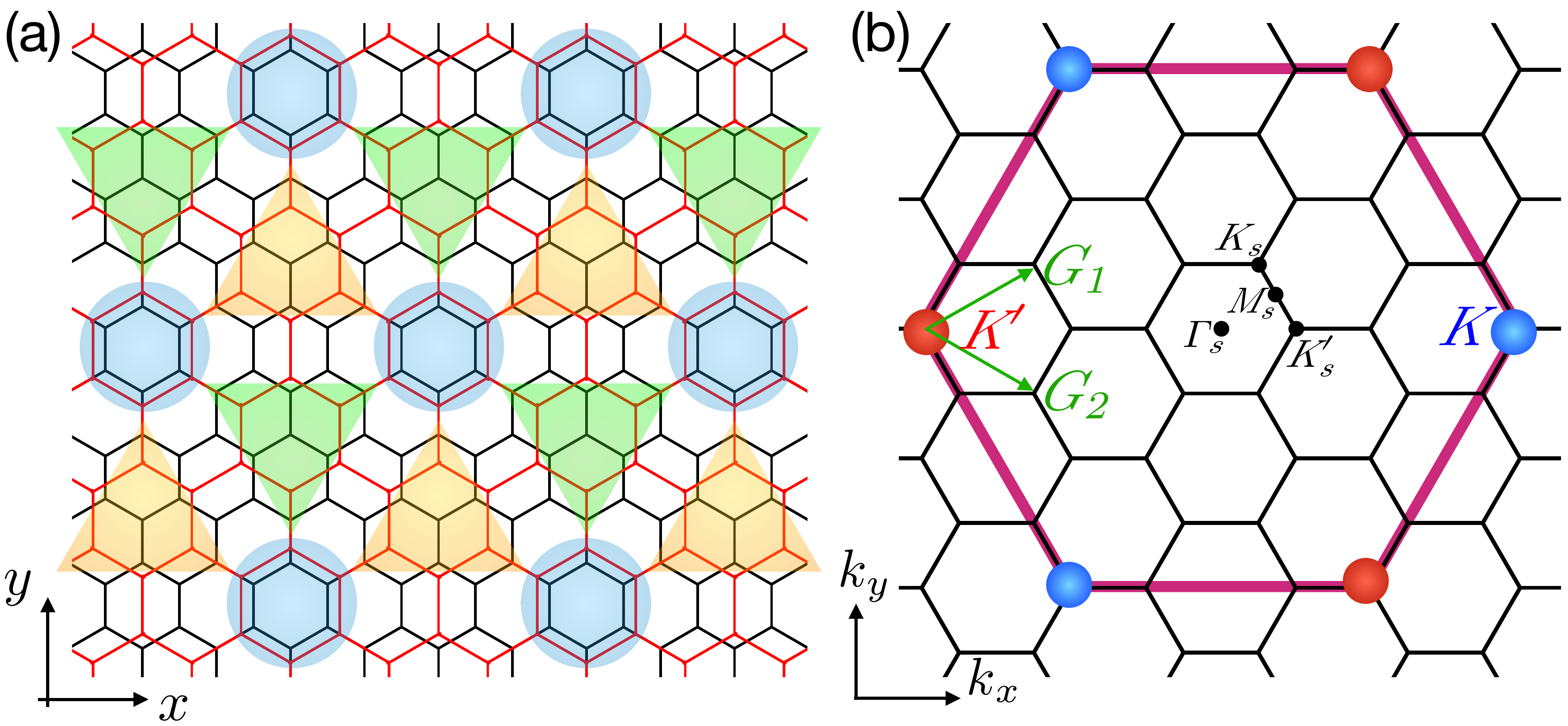}
\caption{Super-lattice structure and Brillouin Zone. (a) Super-lattice
formed by the TLG (black lines) and hBN (red lines). For the sake of
clearness, we exaggerate the lattice constant mismatch to $33\%$. The Moire
pattern is composed of three interlaced regions shaded by blue, yellow, and
green. (b) The Brillouin zone of the TLG on the original lattice (marked by
the purple hexagon) is folded into many mini-Brillouin zone by the Moire
periodic potential. }
\label{lattice}
\end{figure}

For both TLG and hBN, the honeycomb lattice can be bipartitioned into two
triangular sub-lattices. A Dirac cone is generated in the electronic
structure at the charge neutral point (CNP). The Dirac fermions become
massive when the sublattice symmetry $C_{2}\cdot \mathcal{T}$ which relates
the two sub-lattices is broken \cite{Rappe, Miller12PRB}. In the hBN, boron
and nitrogen atoms each form one of the sub-lattices, which breaks the
symmetry between these two sub-lattices. This leads to a large energy gap
(about $2.3$ eV) in the low-lying excitations \cite{Miller12PRB}. In
contrast, the TLG itself is invariant under the sublattice symmetry, which
protects the triple Dirac points near the Brillouin zone (BZ) corners. Thus
the low-energy physics is dominated by the TLG, while hBN just contributes
to a Moire scattering potential under second order perturbation. Such a
Moire potential modulation folds the bands in the original BZ of the
graphene layers into many mini-bands in the mini-Brillouin zones (mBZ), as
displayed in Fig. \ref{lattice}b. As the mBZ is smaller by four order of
magnitude than the original BZ, the bandwidth of the mini-bands is
significantly suppressed. And the mini-bands near the charge neutral point
(CNP) mainly originate from the low-energy valleys ($K$ and $K^{\prime }$
shown in Fig. 1b) in the original TLG \cite{WallbankFalko}.
\begin{figure}[t]
\centering
\includegraphics[width=8.6cm]{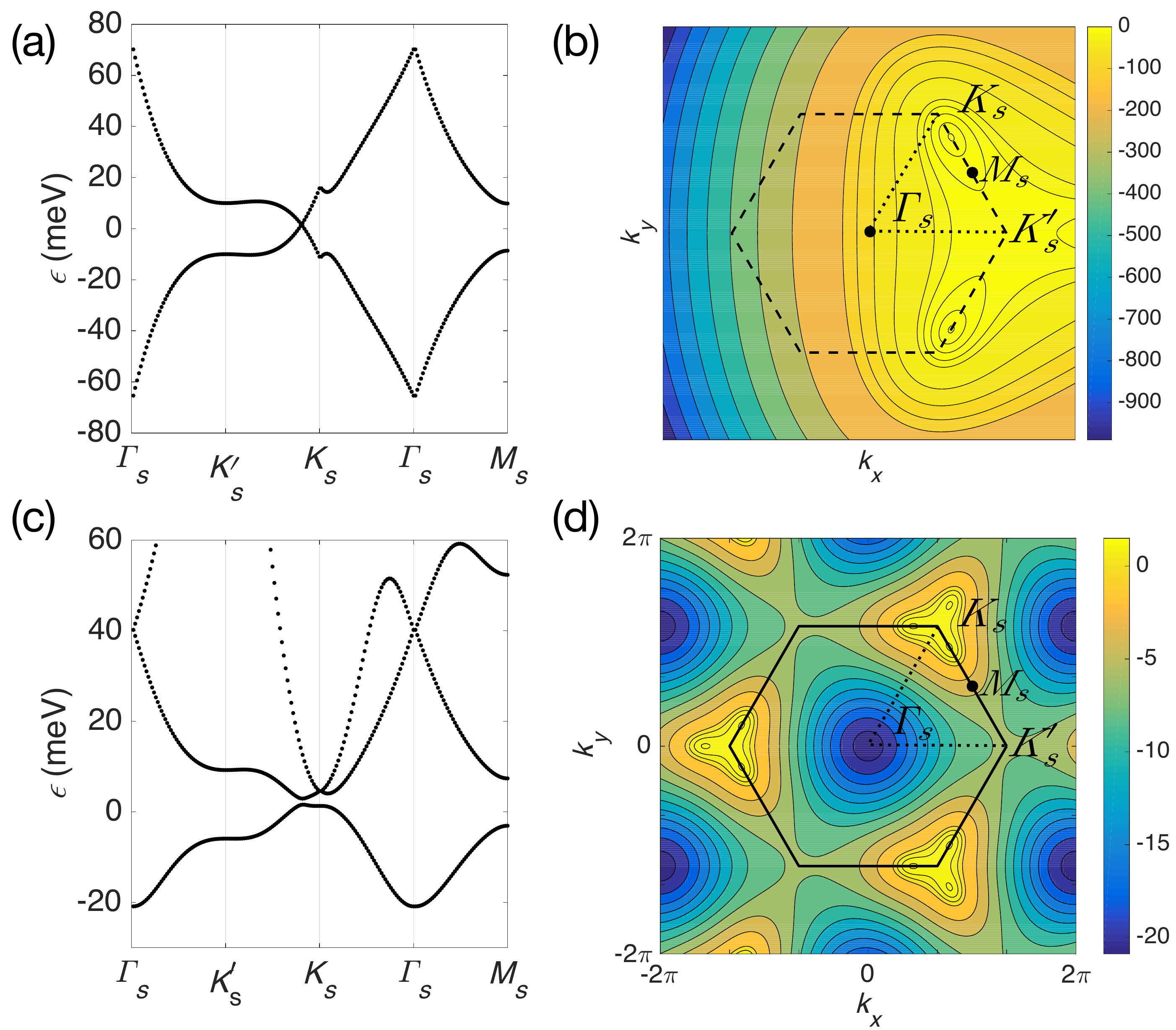}
\caption{Low-energy band structure and contour plot of the valence band. (a)
Low-energy dispersion of the TLG without hBN is displayed in the proximity
of the valley $K$ along the high symmetry lines of the mBZ. (b) Contour plot
of the corresponding valence band near the CNP. The dashed black hexagon
implies the mBZ once hBN is coupled to the TLG. The Dirac points are split
away from $K_{s}^{\prime }$ and extend to almost the vicinity of $K_{s}$.
Triple van Hove points gather near $K_{s}^{\prime }$ instead. (c) Low-energy
Moire band structure for the valley $K$ whose Dirac points are close to $%
K_{s}$ in the mini-BZ. The Dirac points near $K_{s}$ are gapped out by the
Moire potential. In obtaining this band structure, we have adopted the
parameters used in Ref. \protect\cite{KoshinoMcCann} and the Moire potential
strength $80$ meV on the bottom layer of the TLG. (d) Contour plot of the
corresponding valence band near the CNP in the mBZ (black hexagon). The
vicinity of $K_{s}^{\prime }$ hosts three saddle points where the density of
states diverges for this valley band. Color represents energy in unit of
meV. }
\label{bands}
\end{figure}

There are two crucial points about the Moire modulation of the band
structure. First, the two valleys originally connected within one band are
now significantly separated into two degenerate bands, because the valley
distance in the original BZ is significantly longer than the characteristic
wave vector of the Moire potential. This is the reason why the valley degree
of freedom enters into the superlattice as the internal degrees of freedom
of electrons. Second, as the triple Dirac cones are in fact split by
trigonal warping process in the TLG\cite{KoshinoMcCann}, the splitting
distance is relatively small in the original BZ, but quite comparable to the
scale of the folded mBZ (Fig. \ref{bands}a and \ref{bands}b). As a result,
the flat dispersion between the Dirac cones dominate most area of the mBZ,
which further suppresses the kinetic energy. Moreover,  the Dirac
point is gapped out by the interplay between the hBN and TLG, which breaks
the sub-lattice symmetry. A valence band is thus separated from the other
mini-bands by the Moire band gap (Fig. \ref{bands}c), which has four-fold
degenerate associated with the spin and valley degrees of freedom. Electrons
around the valleys $K$ and $K^{\prime }$ are related to each other by either
one of the following transformations: the time-reversal symmetry $\mathcal{T}
$, mirror reflection $M_{x}$, and $C_{6}$ rotation.

Using the effective two-component Hamiltonian for the TLG \cite{KoshinoMcCann},
we have calculated the band structures with the Moire
scattering potential $V_{\rm M}$ assumed to act only on the bottom graphene
layer \cite{FengWang}. Since the two valley bands are connected through the
mirror transformation $M_{x}$, we can just focus on the $K$ valley. In Fig. \ref{bands}c,
the electronic structure for the bands of valley $K$ is
displayed. The contour plot of the corresponding valence band near the CNP
is also shown in Fig. \ref{bands}d. The triple Dirac points originally at
$K_{s}^{\prime }$ are separated to locate along the boundary of the mBZ
towards $K_{s}$, reducing the energy dispersion and inducing the triple van
Hove singularity near $K_{s}^{\prime }$. When the Dirac points are further
gapped out, the remaining triple van Hove points are the most remarkable
fingerprint of the Moire band structure. More precisely, three van Hove
points actually line along the mBZ boundary and center around the zone
corner $K_{s}^{\prime }$. Increasing the value of $V_{M}$ pushes the three
van Hove points towards $K_{s}^{\prime }$. Above all, due to the Moire
scattering and the Dirac physics, the kinetic energy scale is quenched from 1 to 20 meV.
Because the valence band is separated from the other
bands, we are able to write a one-band minimal tight-binding model with
valley and spin degeneracy.

\begin{figure}[t]
\centering
\includegraphics[width=8cm]{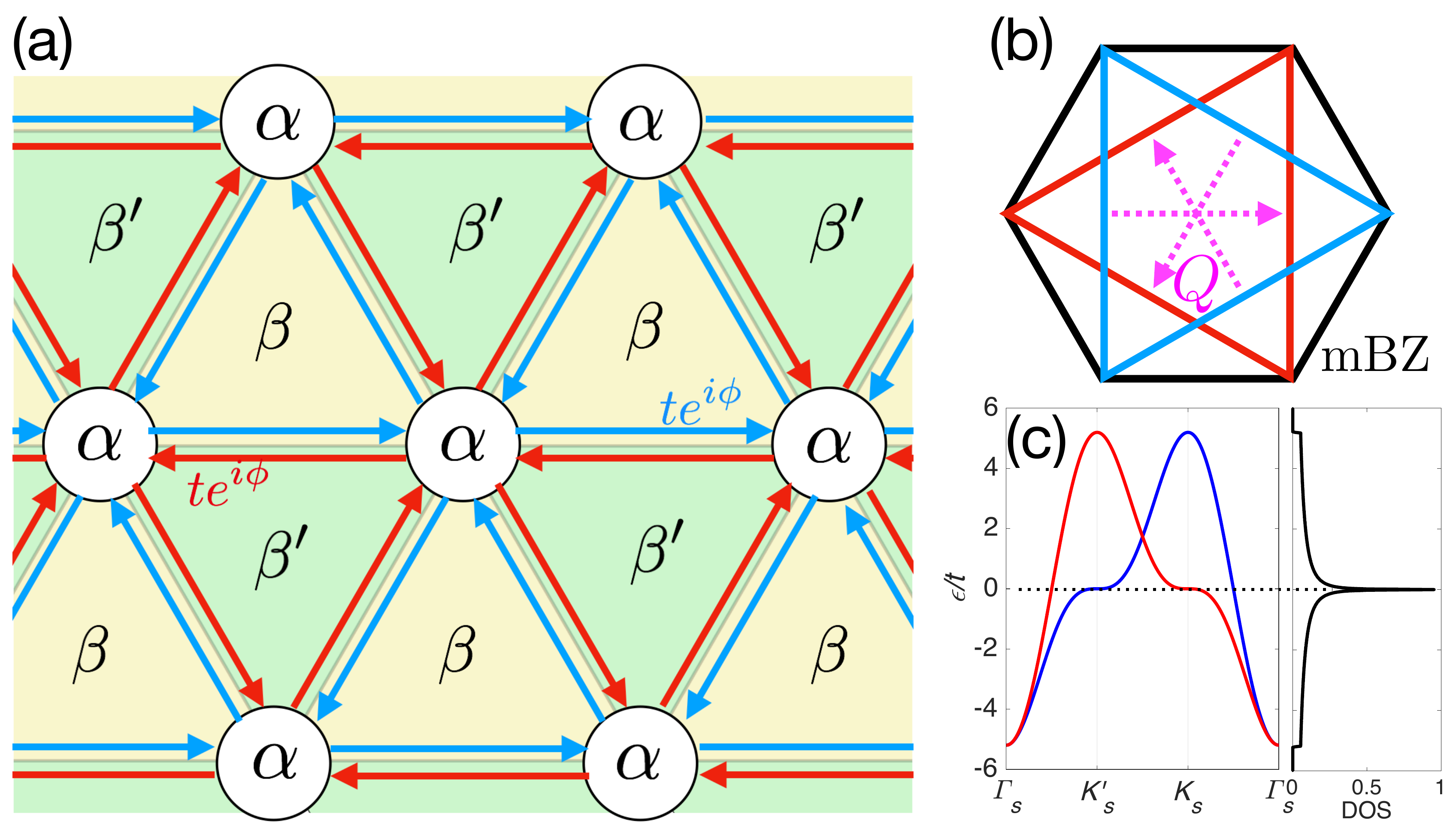}
\caption{Proposed minimal lattice model, Fermi surfaces and their band
structure. (a) The TLG-hBN Moire superlattice composed of three different
zones labelled by $\protect\alpha $, $\protect\beta $ and $\protect\beta %
^{\prime }$, and the pattern of a valley contrasting staggered flux allowed
by $C_{3}$, $M_{x}$, and $T$ symmetries. The $\protect\alpha $ zones form
the effective triangular lattice sites of the Moire superlattice. (b) The
Fermi surfaces of two valleys (the red and blue triangles) at half-filling
are nested by $\mathbf{Q}=(4\protect\pi /3,0)$ and its equivalents connected
by reciprocal unit vectors. (c) Band structures of the two valleys related
by the $M_{x}$-symmetry and the local density of states. }
\label{model}
\end{figure}

Given the Moire mini-band structure, the minimal model should satisfy all
the symmetries mentioned above, and reproduce the key feature of
mini-valence-band: the triple van Hove points and ultra-flat dispersion. In
the triangular Moire lattice sites labelled by $\alpha $ (Fig. \ref{model}a),
one can effectively treat the two valleys as a pair of pseudo-spin denoted
by the Pauli matrices $\tau _{a=x,y,z}$. Since the valley degrees are
decoupled in the band folding, the nearest neighbor hopping should conserve
the valley degrees of freedom. As the two valleys are related to each other
by $M_{x}$ or $\mathcal{T}$, they are intrinsically born of chiral
character. Indeed, the basic symmetries does not forbid the possibility of
chiral flux. However, the time reversal symmetry $T$ requires the two
valleys to have opposite flux phases, and the symmetry $M_{x}$ swaps both
the hopping directions and valley degrees, which fixes the phases in the
hopping integrals (Fig. \ref{model}a). Thus the minimal tight-binding model
for the valence band of the TLG-hBN heterostructure is given by the
Hamiltonian
\begin{equation}
H_{t}=-t\sum_{\mathbf{r},\nu ,\mathbf{\delta }}\left( {\rm e}^{{\rm i}\nu \phi }c_{%
\mathbf{r}+\mathbf{\delta },\nu }^{\dagger }c_{\mathbf{r},\nu }+h.c.\right)
-\mu \sum_{\mathbf{r},\nu }n_{\mathbf{r},\nu },
\end{equation}%
where $\delta =(1,0)$ and $(-1/2,\pm \sqrt{3}/2)$ are the nearest
neighboring vectors of the primitive unit cell, $\nu =\pm $ denote the
valley indices, the fluxes alternate between the $\beta $ and $\beta
^{\prime }$ triangles, and the hopping parameter $t$ measures effectively
the valence bandwidth. For the simplicity, the spin degrees of freedom of
electrons are frozen in our minimal tight-binding model. The flux
penetrating each triangle is given by $\Phi =3\nu \phi $, so this is a
valley-contrasting chiral tight-binding model without breaking the time
reversal symmetry.

In the momentum space, the band dispersion becomes
\begin{eqnarray}
H_{t} &=&\sum_{\mathbf{k},\nu }c_{\mathbf{k},\nu }^{\dagger }\epsilon _{%
\mathbf{k},\nu }c_{\mathbf{k},\nu },  \notag \\
\epsilon _{\mathbf{k},\nu } &=&\varepsilon _{\mathbf{k}}^{e}\cos \phi -\nu
\varepsilon _{\mathbf{k}}^{o}\sin \phi -\mu ,  \notag \\
\varepsilon _{\mathbf{k}}^{e} &=&-2t\left( \cos k_{x}+2\cos \frac{k_{x}}{2}%
\cos \frac{\sqrt{3}k_{y}}{2}\right) ,  \notag \\
\varepsilon _{\mathbf{k}}^{o} &=&-2t\left( \sin k_{x}-2\sin \frac{k_{x}}{2}%
\cos \frac{\sqrt{3}k_{y}}{2}\right) ,
\end{eqnarray}%
where the electronic band dispersion $\epsilon _{k,v}$ generally varies with
the flux phase $\phi $. By observing the band structure with varying $\Phi $%
, we noticed that the flux $\Phi $ essentially tunes those three $C_{3}$%
-related van Hove points. For the valley $K$, when $\Phi $ varies from $0$
to $\pi /2$, these van Hove points approach to $K_{s}^{\prime }$ along the
mBZ boundary. Right at $\Phi =\pi /2$, they merge into one triple van Hove
point, as shown in the expansion around $K_{s}^{\prime }$
\begin{eqnarray}
\epsilon _{\mathbf{k}+\mathbf{K}_{s}^{\prime },+} &=&\left( -6+\frac{3}{2}%
k^{2}\right) \sin \left( \Phi -\frac{\pi }{2}\right) \notag \\
&&+\frac{1}{8}\left( k_{+}^{3}+k_{-}^{3}\right) \cos \left( \Phi -\frac{\pi
}{2}\right) +O\left( k^{4}\right) ,
\end{eqnarray}%
where $k^{2}\equiv k_{x}^{2}+k_{y}^{2}$, and $k_{\pm }\equiv k_{x}\pm {\rm i}k_{y}$%
. When $\Phi $ further increases, the triple van Hove point splits into
three points along the line from $\Gamma _{s}$ to $K_{s}^{\prime }$ and its
equivalents. So the minimal band structure calculated from the low-energy
effective band exhibits that three van Hove points are located right on the
zone boundary in the vicinity of $K_{s}^{\prime }$, and hence the flux of
the minimal model may be slightly smaller than $\pi /2$. In principle, the
exact value of $\Phi $ can be determined by matching the position of the van
Hove points. In the following, we{\ will} focus on the ideal limit $\Phi
=\pi /2$, which reveals the essential correlated physics in the TLG-hBN
heterostructure.

\section{Inter-valley spiral order in the half-filled Mott insulator}

When $\Phi =\pi /2$ at half-filling, the Fermi surface becomes a perfect
triangle that touches the mBZ corners, as shown in Fig. \ref{model}b. In this
case, the two Fermi sheets are perfectly nested and linked by three vectors $%
\mathbf{Q}=(4\pi /3,0)$ and $(-2\pi /3,\pm 2\pi /\sqrt{3})$. However, these
three nesting vectors are equivalent to each other, because they are simply
related by the reciprocal vector of the mBZ. More explicitly, because of the
particle-hole symmetry $\epsilon _{\mathbf{k},+}=-\epsilon _{-\mathbf{k}-%
\mathbf{Q},+}$, we have the relation $\epsilon _{\mathbf{k},+}=-\epsilon _{%
\mathbf{k}+\mathbf{Q},-}$. In such a circumstance, the on-site Coulomb
interactions become important. Taking into account the most relevant on-site
Coulomb repulsion between valleys, we propose a valley version of the
Hubbard model:
\begin{equation}
H_{V}=V\sum_{\mathbf{r}}n_{\mathbf{r},+}n_{\mathbf{r},-},  \label{Eq:int}
\end{equation}%
where $n_{\mathbf{r,}\nu }=\sum_{\sigma }c_{\mathbf{r,}\nu }^{\dagger }c_{%
\mathbf{r,}\nu }$ is the local electron density operator. The perfect Fermi
surface nesting motivates us to introduce the following inter-valley order
parameter
\begin{equation}
\Delta _{\mathbf{Q}}\equiv V\sum_{\mathbf{k}}\langle c_{\mathbf{k}-\mathbf{Q}%
,+}^{\dagger }c_{\mathbf{k},-}\rangle
\end{equation}%
to decouple the Coulomb interaction $V$-term in Eq. (\ref{Eq:int}) as
\begin{equation}
H_{V}\simeq -\sum_{\mathbf{k}}\left( \Delta _{\mathbf{Q}}c_{\mathbf{k}+%
\mathbf{Q},-}^{\dagger }c_{\mathbf{k},+}+h.c.\right) +\frac{\Delta _{\mathbf{%
Q}}^{2}}{V},
\end{equation}%
where $\Delta _{\mathbf{Q}}$ is a spatial uniform order parameter. Under the
mean-field approximation, the above model Hamiltonian can be diagonalized,
and the order parameter is determined by the self-consistent equation
\begin{equation}
\int_{\text{mBZ}}\frac{\sqrt{3}{\rm d}k_{x}{\rm d}k_{y}}{16\pi ^{2}}\frac{V}{\sqrt{%
\epsilon _{\mathbf{k},+}^{2}+\Delta _{\mathbf{Q}}^{2}}}=1,
\end{equation}%
which is similar to the BCS gap equation. If we further assume that the
overall mBZ contribution is dominated by a narrow shell of width $D$ around
the Fermi energy, the solution to the above equation is then given
\begin{equation}
\Delta _{\mathbf{Q}}\simeq De^{-\frac{1}{VN(0)}},
\end{equation}%
where $N(0)$ is the density of states at the Fermi level. At half-filling, $%
N(0)$ diverges, and an infinitesimal interaction $V$ can induce a finite
inter-valley long-range order and gap out the Fermi surfaces completely.
This has been confirmed by the numerical solution to the self-consistent
equation, as shown in Fig. 4a. Actually this is a very peculiar insulating
state induced by the inter-valley scattering $V$. In real space, $\Delta _{%
\mathbf{Q}}$ describes an inter-valley spiral long-range order of the valley
pseudo-spin:
\begin{eqnarray}
\langle \psi _{\mathbf{r}}^{\dagger }\tau _{x}\psi _{\mathbf{r}}\rangle &=&%
\frac{2\Delta _{\mathbf{Q}}}{V}\cos \left( \mathbf{Q\cdot r}\right) ,  \notag
\\
\langle \psi _{\mathbf{r}}^{\dagger }\tau _{y}\psi _{\mathbf{r}}\rangle &=&-%
\frac{2\Delta _{\mathbf{Q}}}{V}\sin \left( \mathbf{Q\cdot r}\right) ,  \notag
\\
\langle \psi _{\mathbf{r}}^{\dagger }\tau _{z}\psi _{\mathbf{r}}\rangle &=&0,
\end{eqnarray}%
with $\psi _{\mathbf{r}}^{\dagger }=(c_{\mathbf{r},+}^{\dagger },c_{\mathbf{r%
},-}^{\dagger })$, and the corresponding configuration is displayed in
Fig. 4b. Therefore, it is this inter-valley spiral phase that describes the
Mott insulating phase observed by the experiment in the TLG-hBN
heterostructure \cite{FengWang}.
\begin{figure}[t]
\centering
\includegraphics[width=8cm]{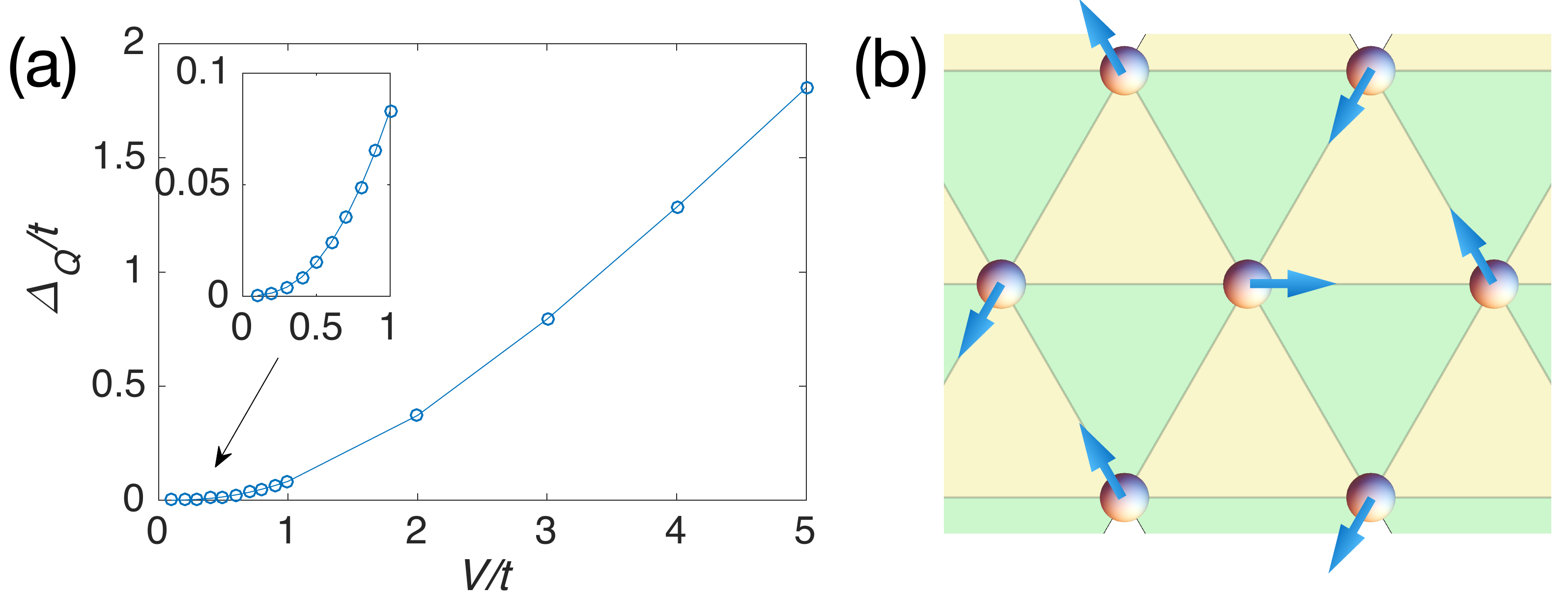}
\caption{The mean-field order parameter and its real space configuration.
(a) Mean-field solution of the inter-valley spiral valley order parameter as
a function of the on-site inter-valley interaction. (b) The real space
configuration of the inter-valley-spiral order. }
\label{order}
\end{figure}

The nesting effect between two valley degrees of freedom successfully
explains the Mott insulating phase at half-filling \cite{FengWang}.
Nevertheless, when the FS shows less prominent nesting, the weak coupling
theory alone can hardly be justified. To probe the complete correlated
physics, we will keep the band structure unchanged to explore the effective
model Hamiltonian in the strong coupling limit.

\section{Mott insulating phase in the strong coupling limit}

Since the strong and weak coupling limits of the conventional one-band
Hubbard model are adiabatically connected to each other, it can be expected
that the correlated physics in the strong coupling regime is closely related
with the weak coupling physics. By assuming $V\gg t$, we can treat the
kinetic term as a perturbation. At zeroth order, the Hilbert space is
separated into two Hubbard-like sub-bands by a charge Mott gap $\sim V$ at
half filling. To pin down the many-body ground state, we need to resort to
the second order perturbation. When the kinetic energy is regarded as a
perturbation, the second order perturbation calculation leads to
\begin{equation}
H_{J}=\emph{P}_{1}H_{t}\frac{1-P}{-V}H_{t}\emph{P}_{1},
\end{equation}%
where $\emph{P}_{1}$ restricts the local Hilbert space to that of one
electron per lattice site $n=1$. The virtual hopping process between the
nearest neighbor sites can induce an antiferro-valley exchange interaction:
\begin{equation}
H_{J}=J\sum_{r,\delta }\mathbf{T}_{r}\cdot \left( {\rm e}^{{\rm i}2\phi T^{z}}\mathbf{T}%
{\rm e}^{-{\rm i}2\phi T^{z}}\right) _{r+\delta },
\end{equation}%
where $J=4t^{2}/V$ and the valley-pseudo-spin operators have been expressed
by
\begin{equation}
T_{r}^{a}\equiv \frac{1}{2}\sum_{\nu ,\nu ^{\prime }}c_{r,\nu }^{\dagger
}\tau _{\nu ,\nu ^{\prime }}^{a}c_{r,\nu ^{\prime }},a=x,y,z.
\end{equation}%
Such an exotic valley-exchange interaction arises from the virtual hopping
process between neighboring sites, which inherits the $\text{SU}(2)_{v}$%
-breaking valley-contrasting flux.

It is known that, for the antiferromagnetic spin-1/2 Heisenberg exchange
interaction on a triangular lattice, the Neel order along $S^{z}$ direction
is frustrated by the lattice geometry and hence the spin moments are
compromised to form a coplanar 120$^{\circ }$ order. The antiferro-valley
exchange interaction, however, carries a flux that can further lower the
ground-state energy for the coplanar order. Apparently, as $\phi =\pi /6$,
each valley pseudo-spin in $H_{J}$ is rotated by $60^{\circ }$ before making
the inner product with its neighbors, and hence every bond wins the most
energy gain $-J/4$ from the $120^{\circ }$ order. From another perspective,
it is such a surprising coincidence that the flux $\phi =\pi /6$ not only
yields the qualitatively correct band structure to match the weak coupling
theory, but also optimizes the ground state energy of the $120^{\circ }$
valley ordering Mott state from the strong coupling limit. It is remarkable
to notice that this state is indeed adiabatically equivalent to the
inter-valley spiral order at half-filling, when the charge degrees of
freedom are frozen by the strong Coulomb repulsion. Both the weak and strong
coupling theories thus point to the same Mott insulating state.

In the above analysis, we have neglected the spin degrees of freedom in our
model formulation for the sake of clearness, because the singular Fermi
surface structure strongly enhance the valley interaction only. When the
spin degrees of freedom is retrieved, however, the corresponding insulating
phase at quarter-filling is given by the spin-polarized antiferro-valley $%
120^{\circ }$ ordering state, while the half-filling insulating phase will
be replaced by the spin-singlet antiferro-valley $120^{\circ }$ ordering
state.

\section{Discussion and Conclusion}

Compared with the magic-angle twisted bilayer graphene \cite{Cao2018Corr},
the kinetic energy of both systems is suppressed by the band folding,
resulting in a similar Moire super-lattice and the mBZ. The Dirac cones of
the twisted bilayer graphene are separated and hybridized, yielding the van
Hove singularity at $M_{s}$ point and flat dispersion in between. Similarly,
in the TLG-hBN, the triple Dirac cones from the ABC stacked trilayer are
separated by the trigonal warping with a strong hybridization, yielding the
triple van Hove points and flat dispersion in between. The drastic
distinction between these two systems is reflected in their symmetries and
Fermi surface structures of the Moire bands. More precisely, with respect to
the same Moire triangular super-lattice, the TLG-hBN is invariant under the
symmetry $M_{x}$ while the twisted bilayer preserves the symmetry $%
M_{y}M_{z} $ instead. Consequently, the $C_{3}$-symmetric Fermi surfaces in
the latter are distinct from that of the former by $30$ degree rotation \cite%
{Cao2016PRL}. If the Fermi surfaces have a nesting effect, the three nesting
vectors are no longer connected by the reciprocal lattice vector. Then the
twisted bilayer graphene will be subjected to an inter-valley triple-$%
\mathbf{Q}$ nesting, and the inter-valley Coulomb repulsion induces a
drastically distinct Mott insulating phase.

What experimental signature could be observed for this order? In the large
length scale of Moire super-lattice characteristic of $\lambda _{M}=15\text{
nm}$, where the valley degrees of freedom are treated as the internal
degrees of freedom inside each Moire supercell, the order with long-range
wave vector $\mathbf{Q}$ exhibits an in-plane spiral feature of the valley
pseudo-spins, but does not involve a density spatial modulation running over
the Moire superlattice. Nevertheless, when zooming into the small
length-scale of the original graphene lattice characteristic of $a=0.246$
nm, the valley degrees of freedom retain their orbital character. The
nesting between the two valleys in the original BZ gives rise to a short
range nesting wave vector, corresponding to a $\sqrt{3}a\times \sqrt{3}a$
charge modulation pattern in the microscopic lattice.

Moreover, in the magic-angle twisted bilayer graphene, unconventional
superconductivity was also observed slightly away from the half-filling \cite%
{Cao2018SC}. Naturally, one would ask whether this TLG-hBN heterostructure
could also become a superconductor by doping away from the half-filling. If
yes, what is the most probable pairing symmetry. From our above analysis,
the inter-valley scattering should still be the most important channel of
pairing interactions, because the intra-valley pairing is not energetically
favored due to the peculiar Fermi surface structures. If only the
inter-valley Coulomb repulsion is considered, there is no privilege between
spin singlet and spin triplet pairing. However, the inter-valley Hund's rule
coupling favors a spin-triplet pairing state. Therefore, the superconducting
state in the TLG-hBN is expected to be in the inter-valley spin-triplet
pairing channel. A detailed discussion on this will be given in a separate
paper.

In conclusion, we have proposed a minimal tight-binding model to describe
the low-energy states of the TLG-hBN super-lattice. Compared to the
low-energy effective bands, the valley-contrasting staggered flux of each
Moire triangle acquires the value of $\pi /2$. At half-filling, the Fermi
surfaces are perfectly nested between the two valleys. This leads to a
strong inter-valley scattering and the system becomes unstable against an
inter-valley spiral order. We believe that this inter-valley spiral ordered
phase is just the Mott insulating phase observed in the experiments \cite%
{FengWang}.

Note added: while in the preparation of this work, we noticed that two
preprints \cite{Senthil,FuLiang} on the model for magic-angle twisted
bilayer graphene appear. One of them \cite{Senthil} proposed a similar
tight-binding model for the TLG-hBN heterostructure.

\textit{Conflict of interest }The authors declare that they have no conflict of interest.

\textit{Acknowledgments: }This work was supported by the National Key
Research and Development Program of MOST of China (2017YFA0302900) and the
National Natural Science Foundation of China (11474331).

\section{Appendix}

The low-energy band structure of the TLG is composed of A-sublattice on the
bottom layer ($A_{1}$) and B-sublattice on the top layer ($B_{3}$), while
the other sublattices are bonded by the on-site interlayer coupling $\gamma
_{1}=0.39\text{ eV}$ and belong to the high-energy sector \cite%
{KoshinoMcCann,MacDonald10ABC}. Therefore, a low-energy effective
Hamiltonian on the two-layer triangular lattice that accounts for the TLG:
\begin{equation}
\hat{H}_{\text{ABC}}^{\text{eff}}=\int d^{2}k\psi _{\mathbf{k}}^{\dagger }H_{%
\text{TD}}(\mathbf{k})\psi _{\mathbf{k}},
\end{equation}%
where $\psi _{\mathbf{k}}\equiv (A_{1,\mathbf{k}},B_{3,\mathbf{k}})^{T}$ is
a two-component spinor. The warped triple Dirac Hamiltonian matrix is given
by
\begin{eqnarray}
H_{\text{TD}}(\mathbf{k}) &=&\frac{v_{0}^{3}}{\gamma _{1}^{2}}\left(
\begin{array}{cc}
0 & \left( \pi ^{\dagger }\right) ^{3} \\
\pi ^{3} & 0%
\end{array}%
\right) +\frac{2v_{0}v_{4}k^{2}}{\gamma _{1}}\left(
\begin{array}{cc}
1 & 0 \\
0 & 1%
\end{array}%
\right)  \notag \\
&&-\left( \frac{2v_{0}v_{3}k^{2}}{\gamma _{1}}-\frac{\gamma _{2}}{2}\right)
\left(
\begin{array}{cc}
0 & 1 \\
1 & 0%
\end{array}%
\right) ,
\end{eqnarray}%
where $\pi =k_{x}+ik_{y}$ and $v_{i}=\sqrt{3}\gamma _{i}a/2$ ($i=0,1,2,3,4$%
). The triple Dirac dispersion is inherited from the three layers of
graphene. On the other hand, the hBN has a large atomic energy $0.8\pm 2.5$
eV, and can therefore be integrated out, leaving a Moire potential
contribution to the TLG \cite{Miller12PRB}:
\begin{equation}
\hat{H}_{\text{Moire}}=V_{M}\sum_{r,G}e^{iG\cdot r}B_{3,r}^{\dagger
}B_{3,r}=V_{M}\sum_{\mathbf{k},G}B_{3,\mathbf{k}-G}^{\dagger }B_{3,\mathbf{k}%
},
\end{equation}%
where $G=\frac{4\pi }{\sqrt{3}}\left( \sin \frac{\pi j}{3},\cos \frac{\pi j}{%
3}\right) $ with $j=1,2,\text{$\ldots $}6$. In the momentum space, $\hat{H}_{%
\text{Moire}}$ essentially scatters the Dirac cones by the vector $G$,
iteration of which gives rise to the Moire reciprocal lattice. Therefore,
the Bloch Hamiltonian in the mBZ can be obtained as
\begin{eqnarray}
\hat{H}_{eff}(\mathbf{k}) &=&\sum_{\mathbf{q}}\psi _{\mathbf{k-q}}^{\dagger
}H_{\text{TD}}(\mathbf{k-q})\psi _{\mathbf{k-q}}  \notag \\
&&+\sum_{\mathbf{q,}G}\psi _{\mathbf{k-q-G}}^{\dagger }\left(
\begin{array}{cc}
0 & 0 \\
0 & V_{M}%
\end{array}%
\right) \psi _{\mathbf{k-q}},
\end{eqnarray}%
where $\mathbf{q}=mG_{1}+nG_{2}$ denotes the reciprocal lattice site. By
exact diagonalizing $\hat{H}_{eff}(\mathbf{k})$, we obtain the low-energy
mini-band structure of the TLG-hBN heterostructure. The calculation result
in this paper is performed by truncating $-5\leq m\leq 5$ and $-5\leq n\leq
5 $. Besides, we perform the Fourier transform to change the Bloch wave
function to the real space for each momentum $\mathbf{k}$:
\begin{equation}
\Psi _{\mathbf{k}}^{\dagger }=\sum_{\mathbf{q}}\left( u_{\mathbf{q}}^{%
\mathbf{k}}A_{1,\mathbf{k-q}}^{\dagger }+v_{\mathbf{q}}^{\mathbf{k}}B_{3,%
\mathbf{k-q}}^{\dagger }\right) .
\end{equation}%
Then from the probability distribution $\left\vert \sum_{\mathbf{q}}u_{%
\mathbf{q}}^{\mathbf{k}}e^{-i\mathbf{q\cdot }r}\right\vert ^{2}+\left\vert
\sum_{\mathbf{q}}v_{\mathbf{q}}^{\mathbf{k}}e^{-i\mathbf{q}\cdot
r}\right\vert ^{2}$, the summation over the valence band gives rise to the
local density of states (LDOS):
\begin{equation}
\rho (x)=\int_{\text{mBZ}}\frac{\sqrt{3}d^{2}k}{8\pi ^{2}}\left\vert \sum_{%
\mathbf{q}}u_{\mathbf{q}}^{\mathbf{k}}e^{-i\mathbf{q\cdot }r}\right\vert
^{2}+\left\vert \sum_{\mathbf{q}}v_{\mathbf{q}}^{\mathbf{k}}e^{-i\mathbf{q}%
\cdot r}\right\vert ^{2}.
\end{equation}%
\begin{figure}[h]
\centering
\includegraphics[width=8cm]{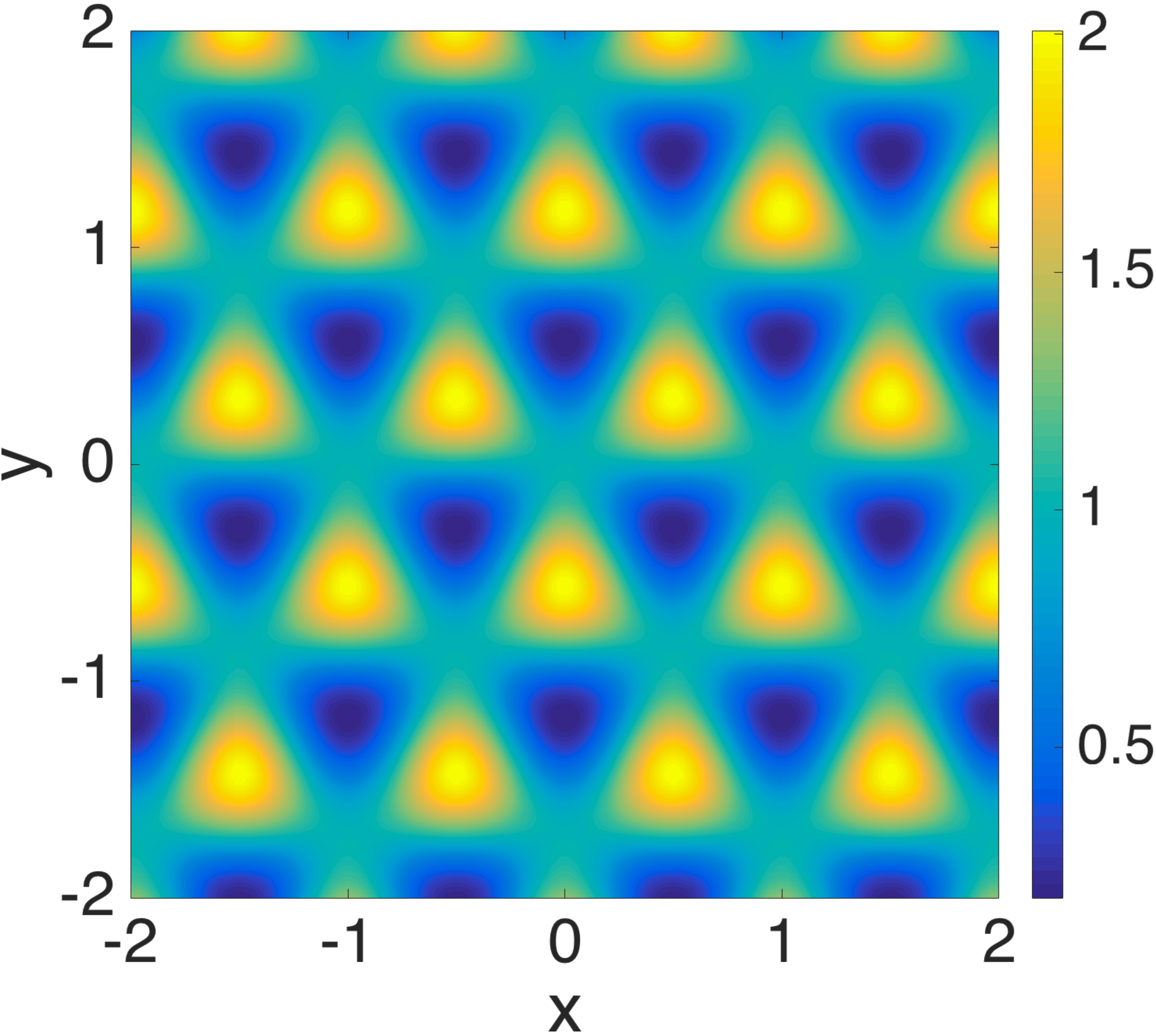}
\caption{The normalized local density of states corresponding to the valence
band. x and y axes are in the unit of Moire wavelength, and the Moire
scattering potential is assumed $V_{M}=80$ meV.}
\label{fig:example}
\end{figure}

Since we are mainly concerned with the valence band, which is well separated
from the conductance band, we do not distinguish the two layer degrees of
freedom. Otherwise there would be a two band model instead. As the two
valley degrees of freedom are related via the mirror symmetry $M_{x}$, they
have the same LDOS distribution, which matches a triangular lattice.
Although the maximum of the LDOS is shifted to $\beta $ zones, the LDOS is
not at all depleted in $\alpha $ zones. The Wannier function is therefore
located on either $\alpha $ or $\beta $ zone. The tight-binding model of the
low-energy valence band should therefore be defined on a triangular lattice
with two valleys on each site. Without the sublattice degrees of freedom, it
leads to an equivalent minimal model whether the Wannier center is located
on the $\alpha $ or $\beta $ zone, as both of them share the same symmetries
$C_{3}$ and $M_{x}$. For the symmetric reason, we have assumed that the
Wannier center is located on the $\alpha $ zones.

\end{document}